\documentclass[final,3p,times,twocolumn]{elsarticle}
\usepackage[hidelinks]{hyperref}
\usepackage{amssymb}
\usepackage{times,latexsym}
\usepackage{url}
\usepackage[T1]{fontenc}
\usepackage{graphicx}
\usepackage{svg}
\usepackage{amsmath}
\usepackage{array}
\usepackage[ruled,vlined,linesnumbered]{algorithm2e}
\journal{Applied Soft Computing}

\begin{document}

\begin{frontmatter}



\title{Towards the generation of hierarchical attack models from cybersecurity vulnerabilities using language models}


\author[inst1]{Kacper Sowka\corref{cor1}}
\ead{sowkak@coventry.ac.uk}
\author[inst2]{Vasile Palade}
\ead{ab5839@coventry.ac.uk}
\author[inst2]{Xiaorui Jiang}
\ead{ad5820@coventry.ac.uk}
\author[inst1]{Hesam Jadidbonab}
\ead{ad4953@coventry.ac.uk}
\cortext[cor1]{Corresponding Author}
\affiliation[inst1]{organization={Centre for Future Transport and Cities, Coventry University},%Department and Organization 
            city={Coventry},
            country={United Kingdom}}

\affiliation[inst2]{organization={Centre for Computational Sciences and Mathematical Modeling, Coventry University},%Department and Organization 
            city={Coventry},
            country={United Kingdom}}

\begin{abstract}
This paper investigates the use of a pre-trained language model and siamese network to discern sibling relationships between text-based cybersecurity vulnerability data. The ultimate purpose of the approach presented in this paper is towards the construction of hierarchical attack models based on a set of text descriptions characterising potential/observed vulnerabilities in a given system. Due to the nature of the data, and the uncertainty sensitive environment in which the problem is presented, a practically oriented soft computing approach is necessary. Therefore, a key focus of this work is to investigate practical questions surrounding the reliability of predicted links towards the construction of such models, to which end conceptual and practical challenges and solutions associated with the proposed approach are outlined, such as dataset complexity and stability of predictions. Accordingly, the contributions of this paper focus on producing neural networks using a pre-trained language model for predicting sibling relationships between cybersecurity vulnerabilities, then outlining how to apply this capability towards the generation of hierarchical attack models. In addition, two data sampling mechanisms for tackling data complexity, and a consensus mechanism for reducing the amount of false positive predictions are outlined. Each of these approaches is compared and contrasted using empirical results from three sets of cybersecurity data to determine their effectiveness.
\end{abstract}


% \begin{highlights}
% \item Natural Language Processing and siamese networks employed to predict relations between public cybersecurity data entries (CVE)
% \item Mathematical methods utilised to curb data imbalance and the effect of anomalous predictions
% \item Capability to predict sibling-level relationships in hierarchy used towards the generation of attack models similar to attack tree
% \end{highlights}


\end{frontmatter}


\section{Introduction}

Within computer-based systems, the emergence of cybersecurity vulnerabilities remains a key concern, as their presence may allow attackers unwarranted control over critical functionality and access to sensitive data. However, given the sheer scale and complexity of modern day systems, it is impossible to anticipate every potential vulnerability and it is thus infeasible to completely secure any system against an attack. Many approaches for dealing with this intractability exist, ranging from security oriented development life-cycles, which aim to minimise the severity and likelihood of emergent vulnerabilities \cite{Felderer2016}, to strategies for effectively responding to a newly discovered vulnerability in the event that it is not discovered prior to release. In the latter case, when cybersecurity experts discover a vulnerability in a system, it is customary to inform the developer and then publish information on the vulnerability publicly, in order to aid the mitigation of future cyberattacks and allow a response (such as a security update) to be mounted. One of the most popular mediums through which this is done is the Common Vulnerabilities and Exposures (CVE) database maintained by the MITRE corporation. As a primarily community run effort, CVE entries are mainly characterised by text descriptions written by volunteers in an informal style describing the nature and occurrence of the given vulnerability.

\begin{figure}[ht]
  \centering
  \includegraphics{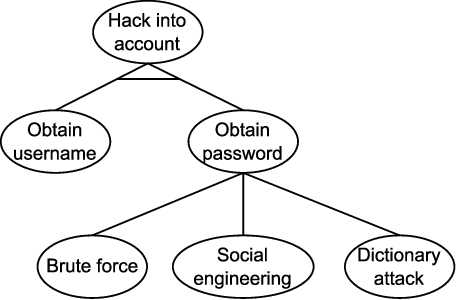}
  \caption{Example attack tree. AND nodes are represented with lines bisecting edges between the parent and children, with OR nodes utilising standard straight lines as edges.}
    \label{fig:at}
\end{figure}

Although there are a wide variety of approaches applied throughout the lifecycle of a software or hardware system, one common approach is the utilisation of models for assessing cybersecurity risks \cite{Kordy2014,pekaric2023streamlining} or enumerating test cases \cite{Felderer2016,Cheah2018}. In particular, attack models are often used as a graphical representation of possible attacks within a given system, which can enable cybersecurity analysts to more intuitively determine areas of high risk. Within this paper, a specific type of attack model, designated here as "hierarchical attack models", are considered as a general class of attack models focused on building a hierarchy grouping vulnerabilities into increasingly high level abstractions. A prevailing example of this is the attack tree \cite{Schneider2000}, which breaks down high-level attack goals, such as "compromise user account", into sub-goals governed by a logical operator, such as "obtain username AND obtain password", and finally terminates at leaf nodes representing atomic attack actions, such as those targeting a specific vulnerability in the system like "brute force attack OR dictionary attack" against a system which fails to restrict the amount of attempts one can make at entering a password. This example can be seen represented graphically in Figure \ref{fig:at}.

In considering how to leverage the incredibly large CVE dataset (with over $200,000$ entries) for the construction of such hierarchical attack models, various key characteristics of the data and the practical utility of attack models must be considered. For one, prior work has demonstrated the feasibility of using public MITRE datasets like CVE as the basis for the generation of hierarchical attack models \cite{pekaric2023streamlining,Falco2018,Jhawar2018}, although existing approaches focus on utilising already existing links within the data \cite{Sowka2023}, which does not offer a feasible approach towards responding to newly discovered vulnerabilities. In addition, due to the size of the CVE dataset and that each entry is written by a 3rd party without major constraints in format or lexicon, it can be challenging for both human experts and algorithmic solutions to effectively discern the interaction of one vulnerability with another. In practical terms, this high volume of entries coupled with a lack of consistent language limits the applicability of simple keyword lookup solutions to determine the likely relationships between CVEs, and introduces a significant amount of complexity when an exhaustive set of combinations for all vulnerabilities is enumerated.

This paper will demonstrate how to neural networks can be utilised for predicting the likelihood of a relationship between vulnerabilities from a public vulnerability dataset like CVE using pre-trained language models. Then, the question of how this capability can be used for grouping vulnerabilities together and lead to the practically viable generation of hierarchical attack models is tackled. In addition, two practical questions relating to data imbalance, and the reliability of using predicted links for generating these groups are tackled, with solutions being demonstrated and empirically verified. Primarily, the focus of this paper is to investigate the ability of these networks to reliably determine the relationships between unseen combinations of known vulnerabilities, and as a further consideration the ability to relate known vulnerabilities to previously unseen (and hence, new) vulnerabilities will also be evaluated. 

In contrast to the traditional formal methods of attack model generation, which requires explicitly defined relationships within the data as a pre-requisite \cite{Sowka2023,Konsta2024}, this paper fundamentally re-frames the problem of generating hierarchical attack models into a data-driven machine learning task and thus paves the way towards full machine learning supported generation, which allows for emergent vulnerabilities to be accounted for. 

Hence, the contributions of this paper are: a) the development of neural networks employing a pre-trained language model for predicting sibling-level relationships between free-form text descriptions of cybersecurity vulnerabilities, b) two data sampling mechanisms to tackle extreme data imbalance for effective training,  c) a consensus mechanism for reducing the impact of false positives in predictions, d) an algorithm for creating basic groups of vulnerabilities based on predicted relationships as a means towards hierarchical attack model generation.

\section{Background and related work}
\subsection{Cybersecurity datasets}

There exists a notable selection of cybersecurity databases, each offering different types of data with different contextualising information at different levels of abstraction, in terms of the types of cybersecurity concept each dataset describes. These include, from most tangible to most abstract: ExploitDB\footnote{https://www.exploit-db.com/}, Common Vulnerabilities and Exposures (CVE) \footnote{https://www.cve.org/}, Common Weakness Enumeration (CWE) \footnote{https://cwe.mitre.org/}, Common Attack Pattern Enumerations and Classifications (CAPEC)\footnote{https://capec.mitre.org/} and the ATT\&CK matrix\footnote{https://attack.mitre.org/}. While ExploitDB considers code-level exploits, CVE has a slightly higher level of abstraction by considering text descriptions of individual vulnerabilities, CWE stores weaknesses that can result in vulnerabilities emerging, CAPEC stores attack patterns which utilise weaknesses and finally ATT\&CK considers the highest level attack goals via tactics and techniques.

A notable characteristic of these datasets are the relationships linking their entries within and between each dataset. For instance, a relatively high level entry from CWE is "CWE-284: Improper Access Control", which has "child" CWEs associated with it, such as "CWE-285: Improper Authorization", which represents a more specific instantiation of the parent weakness. More importantly for the purposes of this work, entries in CWE can be connected to CVE entries, which results in collections of low level individual vulnerabilities based on a relevant CWE weakness. This relationship is made easier to navigate by BRON \cite{hemberg2021linking}, which structures the various MITRE run datasets (CVE,CWE,CAPEC,ATT\&CK) into a bidirectional graph, making traversal and collection simpler. In addition to its primary function in facilitating threat intelligence, BRON can also enable the collection of training data for various cybersecurity relevant machine learning tasks, as it provides textual and relational data. 

In a more generalised context, CVE can be seen as a template for the kind of dataset that a private corporation might maintain for its own cybersecurity activities. While this would likely contain more salient information than just a text description, the relevance of the solutions presented in this paper to such datasets can be justified by the fact that the manner of encoding each entry into neural embeddings can be adapted to any format of data. For instance, though this paper focuses on embedding natural language using DistilBERT, it is also feasible to embed structured text, images or other forms of discrete data into embeddings, though a full discussion of this is out of scope for this paper. 

In addition, the CVE dataset is continuously maintained and new entries are constantly added, thus, while not the primary focus of the foundational work laid out in this paper, the envisioned use case down the line is for networks to demonstrate the ability to determine relationships between newly added vulnerabilities. This remains true for privately maintained vulnerability datasets, since as more information is gathered and new attacks are discovered, new entries are inevitable and the deployment of a neural network capable of determining the relationship of one vulnerability to others can expedite the analysis of how a new vulnerability affects the security of a given system. A key characteristic of the dataset chosen for this paper is that certain groups of vulnerabilities have many more entries than others. Thus, the ability to generalise in a manner that does not "lose" these smaller groups is also explored as in the real world, different types of vulnerabilities will be more or less common, but all equally important to evaluate.

\subsection{Language models for cybersecurity tasks}

Previous work has shown that using pre-trained language models, such as BERT \cite{Devlin2019}, can be used for predicting relationships between various MITRE cybersecurity datasets such as: CVEs and the ATT\&CK matrix \cite{Ampel2021}, CVEs and CAPEC entries \cite{Kanakogi2021}, CVEs and CWE entries \cite{Das2021}, predicting the exploitability rating of CVEs\footnote{Based on https://www.first.org/cvss/} \cite{Yin2020} and generating CVE descriptions from ExploitDB entries \cite{sun2021generating}. Furthermore, there is work on "fine-tuning" the BERT model to cybersecurity domain tasks \cite{Ameri2021,Das2021,Yin2020}, which aims to improve the performance of downstream tasks by allowing the encoder to learn domain specific language features. One issue with deploying a BERT model is its large amount of parameters, as the complexity of the network makes fine-tuning and application resource intensive. Thus, extensions of the BERT model focusing on reducing the model complexity such as "A lite BERT" (ALBERT) \cite{lan2020albert} and DistilBERT \cite{sanh2020distilbert} have been proposed, which use parameter-reduction techniques and knowledge distillation respectively, to increase the efficiency of the BERT model.

Of particular interest in this paper is the way in which several vulnerabilities (e.g. CVE entries) can be assigned to a group (e.g. a CWE), which offers additional high-level insight into how individual vulnerabilities relate to the bigger picture. While CVE entries consist of only a text description in their original source, with additional information like severity being provided via the National Institute of Standards and Technology managed National Vulnerability Database \footnote{https://nvd.nist.gov/}, CWE entries offer a gateway into relationships with other datasets, background details, applicable languages, common consequences, examples, and more.

Das et al\cite{Das2021} explore the use of a transformer-based encoder with a siamese network for predicting links between CVE and CWE entries. To that end, the authors fine-tune a BERT \cite{Devlin2019} model to produce their "V2W-BERT" framework capable of predicting links between CVEs and CWEs with up to 97\% accuracy for CVE entries classified under commonly used CWEs. In addition, the authors introduce various measures to account for rare or "zero-shot" examples for classifying CVEs. This paper will instead focus on the related but novel task of predicting sibling relationships between individual CVEs based on shared CWE "parents", as a means of investigating the performance of such a method in terms of only low-level vulnerability data, which can be utilised in cases where groups of related vulnerabilities are known but the semantic information on what groups them together (such as a CWE description) is not available. In practice, this situation could be encountered if the vulnerability groups were classified based on empirical data, such as after performing a penetration test, and noting that the following vulnerabilities were used in tandem or produced similar results. Thus, during training and prediction, CWEs act purely as groups of related CVEs for the purposes of this paper. While this may seem like an arbitrary distinction, and many of the ideas introduced in this paper could be utilised with an alternative task, this paper focuses on an unexplored approach towards vulnerability data rather than applying an existing approach in a different context in order to investigate a novel perspective which also has security implications.

Another category of methods considers using system logs directly as a form of data on which to utilise neural networks \cite{Li2023}. For instance, Li et al. \cite{Li2023} utilise a transformer network to encode vector representations of system logs during an attack, then utilise an LSTM network to predict attacks and construct an attack graph. This differs from using public cybersecurity data as it focuses on a very specific system implementation, and thus, while it provides much more salient information, this would not be as generally applicable and requires access to system logs in order to perform. However, methods such as this could find use in attack tree generation schemes based on system traces \cite{Sowka2023}.

\subsection{Generating hierarchical attack models}

A common approach towards managing cybersecurity is using attack models to emulate the behaviour of a given system whilst under cyberattack. This can be approached in various ways, with graphical models proving a popular method across various domains \cite{Hong2012,Kordy2014,Hoffmann2015,Li2023,Konsta2024}. In essence, the premise is that the system and/or possible attacks are formally modelled using techniques like graphical and mathematical representations. As mentioned previously, of particular interest in this paper are hierarchical attack models, which are defined here as a broad class of attack models which are structured in a hierarchy, separating different levels of abstraction between individual attack actions and high-level goals. 

In terms of what is actually being modelled, approaches such as the attack tree \cite{Schneider2000,Kordy2014} focus on just the attack model itself, with no explicit modelling of the system itself beyond mentions to what a particular vulnerability targets. Meanwhile, attack graphs generally focus on modeling the physical system such as network connections \cite{Sheyner2002,Hoffmann2015} more explicitly and including them within the graph itself, such that the behaviour of the system being targeted is modeled. Hong et al \cite{Hong2012} introduce the concept of "Hierarchical Attack Representation Models", not to be confused with the general class of hierarchical attack models, which separate system modelling into the "top level" of a hierarchy and the "bottom level" made up of individual vulnerabilities, thus explicitly blending both approaches.

While generation strategies exist for all the above approaches \cite{Hong2013,Sheyner2002,Li2023,Konsta2024}, the focus of this paper is on the generation of hierarchical models in the style of attack trees, particularly from a "bottom-up" or "vulnerability first" perspective, focusing on the lowest level leaf nodes first. While there is a wide variety of approaches in the literature for generating attack trees, including the use of process calculus to represent communications within a system \cite{Vigo2014}, using a formal enterprise model to derive an attack tree \cite{Ivanova2016}, hierarchies of actions sourced from system models or explicitly defined relationships \cite{Gadyatskaya2017} and using a graph based system model to follow the flow of data in a network and derive attack paths \cite{Kern2021}. All of these focus on formally defined algorithms with pre-defined relationships in the data \cite{Sowka2023,Konsta2024}, which do not incorporate advances in machine learning that can open the way for inferring relationships from vulnerability data directly without the need for explicit models of the underlying system and its relationship to the vulnerabilities.

Attack tree generation methods can utilise different information sources to drive its generation, with the use of system models proving initially popular there has been increasing interest in using more generic attack libraries to aid generation, with "model-free" methods that do not necessitate the use of a system model also being proposed \cite{Sowka2023,Pinchinat2020}. Of these, the work of Falco et al \cite{Falco2018} and Jhawar et al \cite{Jhawar2018} make use of MITRE datasets as a generic attack library. Falco et al utilises various datasets to link high-level goals to increasingly tangible tools and actions by traversing the connections ATT\&CK to CAPEC to CWE to finally CVEs. In practice, this follows a system model and existing links within the dataset to string together a set of attacks, weaknesses and vulnerabilities into an attack tree structure, which then enables an analysis to take place \cite{Falco2018}. Jhawar et al focus on a semi-automatic process of enhancing a manually constructed attack tree with generic attack patterns sourced from the link from CVE data (soruced via the NVD) using NLP to break down the description and augment it with affected platforms via the Common Platform Enumeration associated with it. In the end, this information is used to associate the leaf nodes of an existing attack tree with actions like "exploit CVE-XXXX-XXXX" \cite{Jhawar2018}. While these methods differ in both methodology and goals, they both utilise MITRE data by re-interpreting existing links, though with expert inference on existing being present in the work of Jhawar et al. in particular.

A much more recent work has also explored the generation of attack trees by noting the connection between MITRE datasets and hierarchical attack models \cite{pekaric2023streamlining}. Although similar on its surface and tackling a related problem, the work of Pekaric et al. \cite{pekaric2023streamlining} concerns the bottom up structuring of CVE and CWE data into attack trees using the same underlying links between the datasets that this paper used to source its training data. Following this initial structuring of attack trees (or "fragments" as reffered to in the paper), these are interpreted into logical relationships such as OR,AND, SAND using relationship types within CWE data\footnote{https://cwe.mitre.org/data/reports/chains\_and\_composites.html}. Thus, the goals of this work and the work presented in this paper dovetail one another, since the work of Pekaric et al. \cite{pekaric2023streamlining} depends entirely on the data already available, and the work in this paper aims to provide a means to generate what Pekaric et al. label as tree "fragments" from new vulnerability data, without explicitly labelled relationships.

In the interest of integrating recent advancements, the role of large language models in generating hierarchical attack models should be touched on. Another recent publication has for the first time attempted to utilise machine learning for the generation of attack trees, with the stated desire to explore the feasibility of utilising large language models specifically for the generation of attack trees \cite{Gadyatskaya2023}. However, in contrast to bespoke algorithm based approaches with clearly defined data pipelines, this work concerns asking ChatGPT, a commercially available but privately owned model, to create attack trees. These attack trees are then compared to trees found in the literature using metrics developed in preceding literature for quantifying the relationship between two trees, in order to gauge the ability of this model to provide meaningful attack trees. In general, while tangentially related to the type of method explored in this paper, the conclusion of this work is that currently ChatGPT can not be trusted to provide reliable attack trees, certainly not at scale. This is down to many different reasons, better explored in literature focusing on the use of proprietary large language models, but this comes down to the fact that it is too general-purpose to act in a very specific and highly domain sensitive role. Compounded with its proprietary nature, there aren't many productive ways forward to improve upon this model beyond increasingly byzantine prompt hacking. However, this work provides productive insight in how trees generated with the help of machine learning should be evaluated, and cautions against blindly trusting the predictions yielded from a model trained using machine learning. Thus, a bespoke model designed to be used within the confines of an uncertainty-aware methodology allows for more granularity and control over how these inferences are utilised and interpreted.

Despite the popularity of attack trees in the literature, this paper is focused on the more general hierarchy problem rather than tackling attack tree generation head on, as the solutions presented here are generally applicable, and do not focus on attack tree specific problems such as assigning logical relationships to parent nodes. Indeed, no attack tree data is being considered in this paper, rather the relationship between the CVE and CWE dataset, and further CWE to CAPEC and ATT\&CK data, is being drawn as an analogue for a generic cybersecurity hierarchy which bears similarity to attack tree data \cite{pekaric2023streamlining,Falco2018,Jhawar2018}, and the methods outlined in this paper are applicable directly to attack tree generation if the data required to train the necessary networks is available.

\section{Sibling prediction of CVEs}

In the context of relating CVEs to CWEs, a vital question must be answered on how the emergent CVE to CVE relationships are to be interpreted. Namely, if two CVEs are said to be linked to the same CWE, what does that mean on a practical cybersecurity analysis level? Since CWEs represent high-level categories of weaknesses, while CVEs represent individual low-level vulnerabilities, the relationship from CWE to CVE can be said to resemble a parent-child relation. Thus, a CVE sharing a parent with another CVE can be said to have a sibling relationship to that CVE. Hence, in practical terms, a sibling relationship between CVEs mean that these two vulnerability descriptions roughly describe a similar type of weakness, which can also be re-framed as enumerating different ways in which a cybersecurity weakness can be exploited. While this is a fairly abstract way of interpreting CVE to CVE relationships, its a practically feasible and useful way of understanding a huge swathe of historical vulnerability data. Additionally, being able to group together vulnerabilities targetting based on what "type" of weakness they exploit can enable a cybersecurity analyst to better understand a large collection of vulnerability descriptions.

\begin{figure}[ht]
  \centering
  \includegraphics[width=250px]{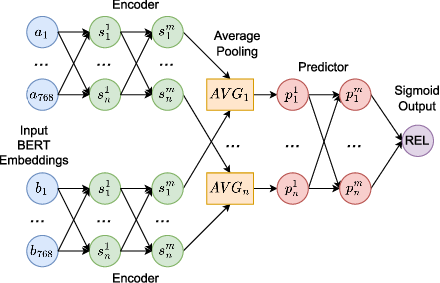}
  \caption{Visualisation of how the prediction network is designed. Encoder and predictor sections have separate hyperparameters, such as the amount of layers (m) and number of neurons (n) in each layer.}
    \label{fig:siamese}
\end{figure}

In terms of implementing a predictive model for sibling relationships, inspired by the approach of Das et al. \cite{Das2021}, a siamese network is employed with "encoder" and "predictor" components separated by an average pooling operation to ensure that the input CVE pairs are order invariant. Figure \ref{fig:siamese} shows the model architecture used in this paper. With the siamese encoder component, input BERT embeddings from vulnerability descriptions are each fed through a feedforward network, which maintains two distinct sets of states for each input while sharing weights, the output of which is then averaged pair-wise to provide a final encoder output. This is then fed to the "predictor" component, which is another feedforward network, ending in a sigmoid output representing a probability distribution over the relationship between the two input vulnerabilities. With the sigmoid output ($REL$ in Figure \ref{fig:siamese}) tending towards $0$ with a negative link and $1$ with a positive link.

In order to train the model, the network requires pairs of embeddings produced by BERT from text descriptions of CVEs alongside their relationship to one another (sibling or not sibling). Given that there are a lot of CVE entries, ($200,000$+), this leads to a substantial amount of potential training data. As each CVE chosen to appear during training needs to be embedded individually by the BERT encoder prior to training, there is a significant pre-processing overhead involved. Therefore, a more compact encoder known as DistilBERT \cite{sanh2020distilbert} will be used in place of the standard BERT model. This is due to both the practical considerations of the experiments performed in this paper, and the wider applicability of the proposed methodology, since it renders the whole process more efficient.

\section{Negative link problem}

\begin{figure}[ht]
    \centering
    \includegraphics[width=150px]{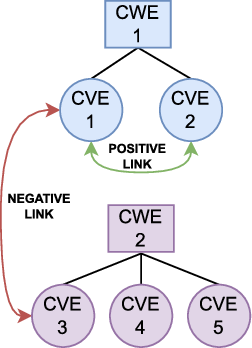}
    \caption{An illustration of the difference between positive and negative links between CVEs.}
    \label{fig:pos_neg_link}
\end{figure}

One factor which must be considered is how positive and negative links are formed, as each CVE will possess a CWE "parent". Notably, the interpretation used by this paper is that if two CVEs share a CWE parent, they are considered related (and thus form a positive link), while any two CVEs which do not share a parent are considered unrelated (negative link). A visualisation of this can be seen in Figure \ref{fig:pos_neg_link}, where fictional CWE 1 has children CVE 1 and 2, while CWE 2 has children CVE 3,4 and 5, with an example positive and negative link being demonstrated.

A major complication with this comes in the mathematical consequences of how positive and negative links are sampled. Let $C=(c_1,c_2,...)$ be a list of cardinalities for a given set of CWEs, denoting the amount of CVEs associated with each CWE such that $c_i=|CWE_i|$. Since positive links are sampled by taking "internal" pairs of nodes, for any given $CWE_i$ the amount of positive pairs acquired is given by $\frac{c_i!}{(c_i-2)! \times 2}$, which can be further simplified to $c_i \times \frac{c_i-1}{2}$. Thus, the total amount of positive links for a given set of node cardinalities $C$ is given by: 
\[
\sum_{i=1}^{|C|} c_i \times \frac{c_i-1}{2}
\]
In contrast, the amount of negative pairs is sourced by the amount of "external" combinations, and thus requires that the cardinalities are multiplied. For instance, the amount of combinations between two sets of CVEs (from two different parent CWEs) is derived by simply multiplying their sizes, so for any given $c_i$ it is necessary to sum the product of $c_i$ with every other member of $C$. This means that the total number of external combinations is given by:
\[
\sum_{i=1}^{|C|}\sum_{j=1}^{|\hat{C}|} c_i \times \hat{c}_j
\]
With $\hat{C}$ denoting $C-\{c_i\}$.

Therefore, negative links will in most cases outnumber positive links, with the imbalance growing proportionally to the amount of CWEs/CVEs. Intuitively, this can be demonstrated by the fact that each time a new entry $c_n$ is added into $C$, it will contribute $c_n \times \frac{c_n-1}{2}$ positive links, which is independent of how many entries there are in the set already, while the amount of negative links will increase by $\sum_{i=1}^{|C|} c_n \times c_i$, which will grow geometrically compared to the linear growth of positive links w.r.t the contents of $C$.

Based on different selections of CVE to CWE links, two techniques for dealing with the negative link problem will be contrasted. 

\subsection{Clique-based sampling}

\begin{figure}[ht]
    \centering
    \includegraphics{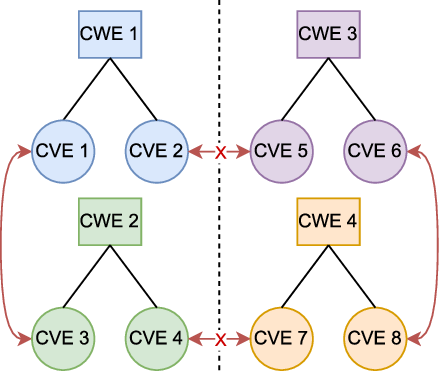}
    \caption{Clique-based sampling of negative links. Red arrows show where negative links are formed with a red "X" crossing an arrow showing a link NOT being formed. The dotted line designates the boundary of two "cliques".}
    \label{fig:clique_eg}
\end{figure}

One approach to mitigating the negative link problem is to define boundaries between cliques of CWEs within which CVEs are permitted to form negative links. This could take the form of clustering together CWEs based on some characteristics, such as the amount of children, or with a specific goal in mind, such as minimising the amount of negative links. However, these are rather arbitrary from the point of view of the data, meaning that there would be a lack of syntactically valid negative links which would likely impact the ability of the model to generalise. Nevertheless, such an approach could be viable in certain domain specific applications, where a carefully curated set of cliques could result in a model designed for tasks in which it is desirable to condition the formation of negative links between only certain groups. Figure \ref{fig:clique_eg} illustrates this idea.

\subsection{Weighted random sampling}

\begin{figure}[ht]
    \centering
    \includegraphics{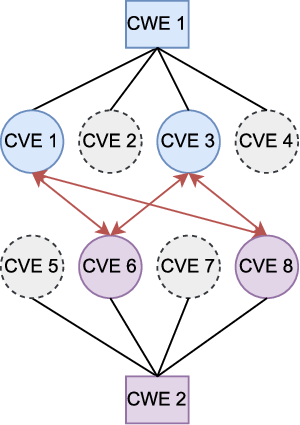}
    \caption{Weighted random sampling of negative links. Greyed-out CVEs with dashed lines are being skipped by the sampling process.}
    \label{fig:random_sampling_eg}
\end{figure}

An alternative solution is to only utilise a subset of all the CVE children to form negative links with, as illustrated in Figure \ref{fig:random_sampling_eg}, based on a weight inversely proportional to the relative size of the CWE. This sampling is controlled by Equation \ref{eq:sampling}, which determines $N_i$ as the percentage of CVEs to sample from the given $CWE_i$.
\begin{align}
    N_i=\left(\frac{\sum(C) -c_i}{\sum(C)} \right)^p
    \label{eq:sampling}
\end{align}
With $c_i$ being the cardinality of the CWE for which $N_i$ is being determined, $\sum(C)$ being the sum of all CWE cardinalities and $p$ being a parameter determining the degree by which larger sets produce smaller samples. Logically, this equation assigns $N$ such that the smaller the cardinality $c_i$ with respect to all of $C$, more of its members will be sampled. This is to ensure that relatively smaller CWEs are not overshadowed by larger ones, and is based on the assumption that over represented CWEs will be less affected by the culling of their negative links. Empirically, the effectiveness of this equation at culling the amount of samples from larger groups while leaving the smaller groups untouched is demonstrated in \ref{sec:appendix_p}. Significantly, Equation \ref{eq:sampling} is not intended to be used to categorically blacklist certain CVEs for the entirety of training, but rather to selectively skip negative links during the sampling of training data such that most CVEs will still end up included at least once in the data.

\section{Towards building hierarchical attack models}

\subsection{Algorithm for grouping vulnerabilities}

This section investigates how the neural networks described in previous sections can be utilised towards producing hierarchical attack models. Firstly, let each pairwise prediction between every vulnerability in a given input set be structured into a "prediction matrix" $P$, where entry $P_{i,j}$ stores the predicted relationship between vulnerabilities associated with indexes $i$ and $j$ respectively. 

\begin{figure}[ht]
    \centering
    \includegraphics{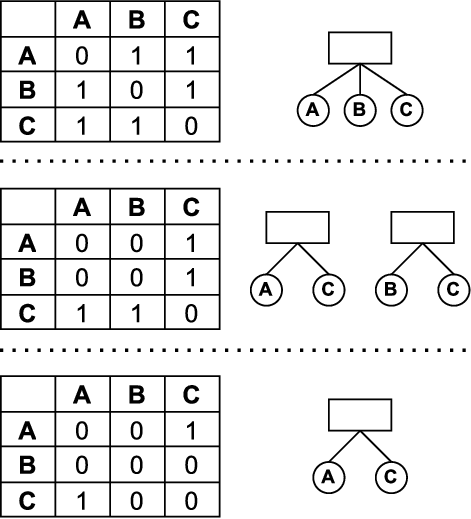}
    \caption{Example of how a prediction matrix can be interpreted into groups.}
    \label{fig:matrix_to_layer}
\end{figure}

Figure \ref{fig:matrix_to_layer} illustrates 3 examples of how a prediction matrix for 3 vulnerabilities $\{A,B,C\}$ can be transformed into consistent groups, with 0 and 1 designating a negative and positive prediction respectively. Algorithm \ref{alg:generation_simple} shows a method for resolving a prediction matrix for an arbitrary amount of vulnerabilities into groups in the manner shown in Figure \ref{fig:matrix_to_layer} by interpreting the problem as a parent-child assignment problem with the prediction matrix acting as a set of constraints for which vulnerabilities can share a parent and which must be separate. To illustrate, in the middle example in Figure \ref{fig:matrix_to_layer} nodes A and B are considered related to C, but not to each other, and so the resultant grouping results in 2 separate copies of C existing in two groups so that A and B do not have to share a group with each other while each getting to share a group with C.

\begin{algorithm}
\SetAlgoLined
\DontPrintSemicolon
\SetKwInOut{Input}{input}\SetKwInOut{Output}{output}
\Input{Prediction matrix ($P$), List of vulnerabilities ($V$)}
\Output{Vulnerability groups based on relationships defined in $P$}
$relationships \leftarrow \emptyset$\;
\For{$i \in V$}{
 \For{$j \in V$}{
   \uIf{$i \neq j$}{
     $rel \leftarrow P_{i,j}$\;
     $relationships \leftarrow relationships \cup \{(\{i,j\},rel)\}$
   }
 }
}
\For{$(\{i,j\},rel) \in relationships$}{
 $forbidden \leftarrow \emptyset$\;
 \For{$(\{i',j'\},rel') \in relationships$}{
   \uIf{$rel' = 0$ \textbf{and} $i = i'$}{
     $forbidden \leftarrow forbidden + j'$\;
   }
 }
 \For{$parent \in i.parents \cup j.parents$}{
   \uIf{$parent.children \cap forbidden = \emptyset$}{
     $i.parent \leftarrow parent$ \;
     $j.parent \leftarrow parent$ \;
   }
 }
 \uIf{No suitable parent found}{
   $parent \leftarrow$ Create new parent\;
   $i.parent \leftarrow parent$ \;
   $j.parent \leftarrow parent$ \;
}
}
 \KwRet set of groups of vulnerabilities\;
 \caption{Parent assignment with respect to predicted sibling relationships}
 \label{alg:generation_simple}
\end{algorithm}

While not an attack model generator in itself, Algorithm \ref{alg:generation_simple} could be used in a layer-wise manner to produce a model like an attack tree \cite{Kordy2014,Schneider2000,pekaric2023streamlining}, which structures vulnerabilities in a hierarchy based on logical operators as parents (e.g. to enter a user account you must obtain the username AND password). This would require two additional functionalities: a multi class prediction model incorporating logical operators (OR,AND) and a way to incorporate information from the vulnerabilities into upper layers. Meaning, that the sibling prediction network should predict the type of logical relationship connecting two vulnerabilities rather than simply determining if there is one, and there should be some mechanism for the prediction network to understand newly created "parents" alongside the NLP encoded vulnerabilities to allow the top side of the structure to be created mingling both "leaf" and "inner" nodes together.

\subsection{Consensus mechanism}

If we contrast the task being learned by the neural networks with the task that they are being applied to, it can be said that the sibling prediction network is trying to maximise the probability that it correctly classifies the relationship between two vulnerabilities, while the task being performed by Algorithm \ref{alg:generation_simple} is attempting to maximise the probability that the group generated resembles an actual group of vulnerabilities in the underlying data. This is done by treating each pair-wise prediction as categorical constraints on which vulnerabilities can appear together. Thus, false positive and false negative predictions can significantly disrupt the generation process, even if they are incredibly rare, as each negative prediction that strays from the prevailing trend can induce the generation of two incorrect groups in place of one correct grouping (see Figure \ref{fig:matrix_to_layer}). As such, in order to perform sampling in a more holistic manner, the way that these predictions are interpreted should somehow be conditioned by other predicted links, instead of being taken in isolation.

A simple trust model \cite{Braga2019} is proposed here to quantify "evidence" for how likely a predicted link is to be true based on the predictions made for other related vulnerabilities. As a simple model, this will consider only direct links (e.g. siblings will be considered, but not siblings of siblings). As a running example, let $V=(A,B,C,D,E)$ be an ordered list of vulnerabilities, and $P$ the prediction matrix where each entry $P_{i,j}$ indexes predicted relationships according to the order of $V$:
\[P=
\begin{bmatrix}
0 & 1 & 1 & 1 & 1 \\
1 & 0 & 1 & 1 & 0 \\
1 & 1 & 0 & 1 & 0 \\
1 & 1 & 1 & 0 & 0 \\
1 & 0 & 0 & 0 & 0
\end{bmatrix}
\]
$A$,$B$,$C$ and $D$ are all considered siblings, however $A$ is the only entry to consider $E$ a sibling. Intuitively, it can be concluded that due to the other siblings of $A$ all agreeing that $E$ is unrelated to them, it is likely that the prediction that $A$ and $E$ are siblings is a false positive. To allow $A$ to arrive at this conclusion, a trust function must be defined, which quantifies the "opinion" of an intermediary vulnerability on the supposed connection. Thus, in each computation of trust, there are 3 entities of interest: the original, intermediary and suspected vulnerabilities. In essence, the original vulnerability is the one asking the intermediary vulnerability if the suspected vulnerability can be trusted (aka, if its related).
\begin{align}
    f(s,r) = \begin{cases}
      0  & \text{if } s=0 \\
      1 & \text{if } s=1 \text{ and } r=1 \\
      -1 & \text{if } s=1 \text{ and } r=0
    \end{cases}
    \label{eq:trust_func}
\end{align}

Equation \ref{eq:trust_func} defines a simple trust function $f(s,r)$ where $s$ determines if the intermediary vulnerability is a sibling of the original vulnerability and $r$ is the relationship between the intermediary vulnerability and the suspected vulnerability. For example, to quantify the opinion of $B$ on if $A$ and $E$ are related: \[
f(P_{A,B},P_{B,E})=f(P_{0,1},P_{1,4})=f(1,0)=-1
\]
If $B$ was not a sibling of $A$, then its relationship to $E$ would have no bearing on the final consensus, hence $f$ returns 0 for cases where the intermediary vulnerability is unrelated; but since $B$ is a sibling of $A$, and $B$ considers $E$ unrelated, $B$ contributes a $-1$ to the consensus.
Before arriving at a final consensus, a "score" matrix $S$ must first be assembled to quantify the effects of $f$ on the prediction matrix $P$ as a whole, for instance the "evidence" of the link between $A$ and $E$ from its other siblings $C$ and $D$ should also be quantified and tallied. Each element in $S$ is calculated based on $P$ using Equation \ref{eq:score_matrix}.
\begin{align}
    S_{i,j}=P_{i,j}+\sum_{k=1}f(P_{i,k},P_{k,j}) \text{ for } i \neq j \neq k
    \label{eq:score_matrix}
\end{align}

Note that the score matrix is not symmetrical, as $E$ gauges its own opinion on all other vulnerabilities based on its one link to $A$ and thus in isolation will conclude that it too should be related to all these vulnerabilities. Hence, a final "consensus matrix" $C$ can be derived from the score matrix using Equation \ref{eq:consensus_matrix}, which effectively folds the matrix in on itself to normalise all opinions in both directions and convert the predictions into discrete $1$ or $0$ values.
\begin{align}
    C_{i,j}= \begin{cases}
              1 & \text{if } S_{i,j}+S_{j,i} > 0 \\
              0 & \text{otherwise}
            \end{cases}
    \label{eq:consensus_matrix}
\end{align}

To illustrate, take the previous example of $P$ for the vulnerabilities $\{A,B,C,D,E\}$, using Equations \ref{eq:score_matrix} and \ref{eq:consensus_matrix} the following can be derived: \[
S=\begin{bmatrix}
0 & 2 & 2 & 2 & -2 \\
3 & 0 & 3 & 3 & -1 \\
3 & 3 & 0 & 3 & -1 \\
3 & 3 & 3 & 0 & -1 \\
1 & 1 & 1 & 1 & 0
\end{bmatrix}
C=
\begin{bmatrix}
0 & 1 & 1 & 1 & 0 \\
1 & 0 & 1 & 1 & 0 \\
1 & 1 & 0 & 1 & 0 \\
1 & 1 & 1 & 0 & 0 \\
0 & 0 & 0 & 0 & 0
\end{bmatrix}
\]

As can be seen in matrix $C$, the suspected false positive has been removed and now all the sibling relationships are internally consistent. In datasets like the CVE/CWE relationships being used in this paper, this approach can lead to a significant decrease in "anomalous" groups being formed by Algorithm \ref{alg:generation_simple}, as will be shown empirically further on.

\section{Experimental results}

Within this work, the key intention is to explore the ability of neural networks using DistilBERT encoding of vulnerability descriptions to learn to predict previously unseen pairs of vulnerabilities. This means that in the following experiments, the primary objective is to determine if these networks can successfully predict relationships between vulnerabilities that have been seen during training, but with unseen pairings of these known vulnerabilities. An additional dimension to this is the potential for brand new vulnerabilities to be evaluated using these same networks. In short, there are 2 separate levels of generalisation considered here: generalising to unseen links between previously seen vulnerabilities, generalising to unseen links by virtue of showing completely unseen vulnerabilities. Finally, these different abilities to generalise can be shown using 3 experimental environments: testing unseen pairings of known CVEs (old-to-old), testing known CVEs against unknown CVEs (old-to-new), and testing all pairings of unseen CVEs (new-to-new).

Note that at the present, the ability to generalise this method to pairing new CVEs with new CVEs is out of scope, but the eventual end goal of a vulnerability-to-vulnerability prediction scheme is to enable new to new vulnerability prediction. To that end, the following experiments aim to gauge this ability on 3 separate collections of CWEs, with an additional 2 modifications being introduced for the express purpose of evaluating the ability to generalise to unseen CVEs. These data collections have been assembled using the 2 data sampling mechanisms introduced above. Further, these experiments are also designed to take into account different scales, with the amount of CWEs and their respective cardinalities being a key variable, and the intended effect is to observe just how much smaller CWEs get "lost" in the training data. This is done by noting the average, maximum and minimum sizes of CWE groups, and producing a correlation coefficient between the size of the group and its accuracy and F1 score for fine-tuned and base data. Another metric aiming to highlight the performance of low size CWE groups is taking the accuracy and F1 score of each CWE in isolation, then taking an average where every group has equal weight. In practice, this results in the more common low-size CWEs bringing down the average metric to demonstrate their relative poor performance.

Finally, the ability to deploy these networks towards generation of hierarchical attack models is evaluated, on unseen pairings of known CVEs and on a mixture of known and unknown CVEs. In addition, these experiments also demonstrate if the consensus mechanism introduced above can help in increasing the reliability of these predictions.

All training data was collected from BRON on 02/08/2022 and structured into pairs according to the two strategies outlined prior in Sections 4.1 and 4.2 (clique based and weighted random sampling respectively). Next, $80\%$ of all data was used as training data, and 5-fold cross validation was used to optimise hyperparameters to avoid potential bias towards a specific data selection. Networks ultimately trained on the entire training data were tested using a the remaining 20\% of data (consisting of unseen pairings of CVEs). Finally, the ability of the networks to successfully determine the relationship between old CVEs (aka those included in the training data) and new CVEs (released or associated with CWEs after 02/08/2022) is also quantified using a separately acquired dataset\footnote{See \ref{sec:data_appendix} for details}.

Most experiments (with the exception of the consensus mechanism experiments) contrast the results of base DistilBERT encoder with an encoder fine-tuned using Masked Language Modelling, with the fine-tuning process following guidelines set out by Devlin et al\cite{Devlin2019} in the original BERT paper, with more details on the process in \ref{sec:appendix_MLM}. In short, the best performing fine-tuned model was achieved with a batch size of $16$ and a learning rate of $5.00E-05$, the parameters were frozen during sibling prediction training for both fine-tuned and base DistilBERT models. Results tables contain 4 commonly used machine learning metrics: Acc = Accuracy, Pre = Precision, Rec = Recall, F1 = F1 Score. Hyperparameters common to all neural network experiments include: 100 epoch training cycles, categorical cross entropy loss, CLS pooling for DistilBERT, a hidden architecture of $512,256,128,64$ neurons in each layer for the siamese encoder, followed by $128,64,32,16$ for the prediction network. A $0.1$ dropout layer after the average pooling operation (AVG in Figure \ref{fig:siamese}) and $0.1$ L2 regularisation on neurons in the prediction network were also used.

For clique-based data in Dataset 1, negative links were drawn entirely from within each clique, with training pairs from all cliques being mixed together, shuffled and split into a training and test set. For Datasets 2 and 3, pairs were sampled from all CWEs using weighted random sampling, with the training and test set splits being derived from a common, randomly shuffled pool that included all CWEs.

\subsection{Dataset 1}

Dataset 1 is the largest, incorporated 3 cliques of 5 CWEs each, whose breakdown can be seen in \ref{sec:data_appendix}. This set was chosen at random to contain relatively low-cardinality CWEs. With a grand total of $2283$ unique CVEs and a negative to positive ratio of roughly $33:50$. A learning rate of $5e-7$ and batch size of $32$ was used. Overall results for Dataset 1 can be seen in Table \ref{tab:clique_results_2}. 

\begin{table}[h]
    \centering
    \caption{Results of the sibling prediction experiment on Dataset 1}
    \begin{tabular}{|l|l|l|l|l|}
    \hline
         & \textbf{Acc \%} & \textbf{Pre \%} & \textbf{Rec \%} & \textbf{F1 \%} \\ \hline
        \textbf{Old-Old Fine-tune} & 99.99 & 99.99 & 99.98 & 99.99 \\ \hline
        \textbf{Old-Old Base} & 99.38 & 99.94 & 99.04 & 99.48  \\ \hline
        \textbf{Old-New Fine-tune} & 85.87 & 83.85 & 91.45 & 87.49 \\ \hline
        \textbf{Old-New Base} & 79.96 & 80.95 & 82.23 & 81.59  \\ \hline
        \textbf{New-New Fine-tune} & 72.04 & 76.28 & 78.19 & 77.22 \\ \hline
        \textbf{New-New Base} & 66.31 & 77.54 & 62.54 & 69.24  \\ \hline
    \end{tabular}
    \label{tab:clique_results_2}
\end{table}

Taken at a glance, these results do not tell the full story however, as there are 15 CWE groups with varying sizes being flattened into 4 overall metrics. Overall, the 3 cliques had an average CWE cardinality of $151$, with a maximum of $868$ and minimum of $2$. On the basis of results obtained from testing on individual CWE groups, $3$ CWEs had not a single positive example in the test set due to its small size, with another having only a single incorrectly classified positive sample. Using the Pearson correlation coefficient, in the fine-tuned data a positive correlation of $0.29$ can be computed between the size of the CWE group and the accuracy of its prediction, and a much stronger positive correlation of $0.6$ can be drawn between the size of the group and the F1 score. In addition, the base sample had a $0.14$ correlation between size of CWE and accuracy, and a $0.67$ between size and F1 score. When weighed equally, the average accuracy of all CWEs taken individually is $99.84$ for fine tuned and $99.36$ for base, however, the average F1 score is $53.3$ for fine-tuned and $44.34$ for base.

\subsection{Dataset 2}

Details of the composition of Dataset 2 can be found in \ref{sec:data_appendix}. For this set, 8 CWEs and $p=2$ was chosen to achieve a ratio of 2:3 between negative and positive links. This experiment utilised a learning rate of $5e-7$ with a batch size of $32$, the average CWE cardinality was $316.75$, with a maximum of $1091$ and a minimum of $12$. Results for Dataset 2 can be seen in Table \ref{tab:random_results}. This time, interestingly, the Pearson correlation coefficient between accuracy and size of CWE was $-0.77$ for fine-tuned and $-0.84$ for base, for F1 score this was $0.92$ for fine-tuned and base. When weighed equally, the average accuracy between each CWE group was $88.43$ for fine-tuned and $82.72$ for base, with the average F1 score being $30.65$ for fine-tuned and $27.90$ for base.

\begin{table}[h]
    \centering
    \caption{Results of the sibling prediction experiment on Dataset 2.}
    \begin{tabular}{|l|l|l|l|l|}
    \hline
         & \textbf{Acc \%} & \textbf{Pre \%} & \textbf{Rec \%} & \textbf{F1 \%} \\ \hline
        \textbf{Old-Old Fine-tune} & 97.79 & 96.81 & 99.08 & 97.93 \\ \hline
        \textbf{Old-Old Base} & 93.09 & 89.08 & 99.04 & 93.80  \\ \hline
        \textbf{Old-New Fine-tune} & 82.68 & 62.53 & 70.93 & 67.03 \\ \hline
        \textbf{Old-New Base} & 75.00 & 49.76 & 75.45 & 59.97  \\ \hline
        \textbf{New-New Fine-tune} & 77.53 & 51.56 & 46.40 & 48.84 \\ \hline
        \textbf{New-New Base} & 70.56 & 40.46 & 57.97 & 47.66  \\ \hline
    \end{tabular}
    \label{tab:random_results}
\end{table}

\subsection{Dataset 3}

Details of the composition of Dataset 3 can be found in \ref{sec:data_appendix}, the design of this data sample was to include more high cardinality CWEs while maintaining the same hyperparameters. For this set of CWEs the value of $p=1$ was chosen to achieve a better balance of negative and positive links, leading to a ratio of roughly 33:50 negative to positive links. Same hyperparameters were used as in Dataset 2, but with a larger mean CWE cardinality of $573.25$, from a maximum of $2789$ and minimum of $44$. Table \ref{tab:random_results_2} shows the results for Dataset 3. Once again, a negative correlation coefficient between accuracy and size of clique was computed, with $-0.64$ for fine-tuned and $-0.63$ for base. For F1 score, the correlation was $0.65$ for fine-tuned and $0.6$ for base. When taken individually, the average accuracy per CWE was $88.87$ for fine-tuned and $89.49$ for base, with the average F1 score being $36.33$ for fine-tuned and $39.73$ for base.

\begin{table}[h]
    \centering
    \caption{Results of the sibling prediction experiment on Dataset 3.}
    \begin{tabular}{|l|l|l|l|l|}
    \hline
         & \textbf{Acc \%} & \textbf{Pre \%} & \textbf{Rec \%} & \textbf{F1 \%} \\ \hline
        \textbf{Old-Old Fine-tune} & 99.19 & 98.86 & 99.80 & 99.33 \\ \hline
        \textbf{Old-Old Base} & 99.71 & 99.64 & 99.88 & 99.76  \\ \hline
        \textbf{Old-New Fine-tune} & 84.78 & 62.97 & 79.84 & 70.41 \\ \hline
        \textbf{Old-New Base} & 84.98 & 65.88 & 70.04 & 67.90  \\ \hline
        \textbf{New-New Fine-tune} & 49.92 & 27.57 & 55.07 & 36.74 \\ \hline
        \textbf{New-New Base} & 80.81 & 50.12 & 39.02 & 43.88  \\ \hline
    \end{tabular}
    \label{tab:random_results_2}
\end{table}

\subsection{Generation experiment}

In order to gauge the effectiveness of the consensus mechanism, while also demonstrating the ability of the networks trained on weighted random data to generalise to mainly unseen negative links, this experiment concerns groups created using Algorithm \ref{alg:generation_simple} on the test sets of Dataset 2 and 3, with and without using the consensus mechanism outlined in Section 5.2. During prediction, the best performing network was used in each case, meaning that the fine-tuned encoder was used for Dataset 2 predictions, and the base encoder was used for Dataset 3 predictions. 

Firstly, for each Dataset, 100 CVE sets were prepared at random from the two, three and four largest CWEs such that the node selection produced pairings which never appear in training data. Using the largest CWEs means that there is a minimal amount of potential negative links in the training data; while maximising the potential pool of vulnerabilities, making the sampling easier, with each CWE contributing a maximum of 10 CVEs. Following this, a prediction matrix $P$ was prepared for each set using the relevant neural network alongside a consensus matrix $C$ using Equations \ref{eq:score_matrix} and \ref{eq:consensus_matrix}, after which Algorithm \ref{alg:generation_simple} was used to generate groups for both matrices.

Results for this experiment on unseen pairings of old CVEs can be seen in Table \ref{tab:consensus}, with the columns explained as such: "Dataset" is the dataset from which entries were sampled, "\#" being the amount of CWEs included in the sampling process (with each only being allowed to contribute a max of 10 CVEs), "Direct" displaying results using just matrix $P$ and "Consensus" showing results with the consensus mechanism via matrix $C$. This is computed as the average similarity of the generated group to the real CWE group it most resembles. This similarity metric is obtained using the Jaccard similarity coefficient, which was also used by Gadyatskaya et al. for gauging the similarity of the "ground truth" trees in literature to the ones generated by ChatGPT \cite{Gadyatskaya2023}.

\begin{table}[h]
    \centering
    \caption{Results for 100 randomly chosen sets of known CVEs for Dataset 2 and 3, showing the effectiveness of the consensus mechanism.}
    \begin{tabular}{|c|c|c|c|}
    \hline 
        \textbf{Data} & \textbf{\#} & \textbf{Direct} & \textbf{Consensus} \\ \hline
         & \textbf{2} & 69.01 & 88.71 \\ \cline{2-4}
        \textbf{Dataset 2} & \textbf{3} & 71.68 & 91.29 \\ \cline{2-4}
         & \textbf{4} & 65.61 & 78.85 \\ \hline
         & \textbf{2} & 75.33 & 94.14  \\ \cline{2-4}
        \textbf{Dataset 3} & \textbf{3} & 84.82 & 98.34  \\ \cline{2-4}
         & \textbf{4} & 74.84 & 88.25 \\ \hline
    \end{tabular}
    \label{tab:consensus}
\end{table}

In addition, the same experiment was performed for 100 sets of 6 old and 4 new CVEs, this time examining the top 3 largest CWEs against one another, the results of which can be seen in Table \ref{tab:consensus_old_new}.

\begin{table}[h]
    \centering
    \caption{Results for 100 randomly chosen sets of known and unknown CVEs for Dataset 2 and 3.}
    \begin{tabular}{|c|c|c|c|}
    \hline 
        \textbf{Data} & \textbf{CWEs (IDs)} & \textbf{Direct} & \textbf{Consensus} \\ \hline
        & \textbf{74, 326} & 52.5 & 70.23 \\ \cline{2-4}
        \textbf{Dataset 2} & \textbf{326, 400} & 52.93 & 73.71 \\ \cline{2-4}
        & \textbf{74, 400}& 62.97 & 77.5 \\ \cline{2-4}
        & \textbf{74, 326, 400} & 48.97 & 62.44 \\ \hline
        & \textbf{203, 94} & 34.19 & 37.27 \\ \cline{2-4}
        \textbf{Dataset 3} & \textbf{94, 306} & 48.27 & 60.1 \\ \cline{2-4}
        & \textbf{203, 306} & 57.58 & 78.61 \\ \cline{2-4}
        & \textbf{203, 94, 306} & 50.91 & 67.34 \\ \hline
    \end{tabular}
    \label{tab:consensus_old_new}
\end{table}

\section{Discussion \& future work} 

\subsection{Discussion of empirical results}

\subsubsection{Performance of neural networks}

It can be clearly seen that as long as both CVEs have been previously seen in some form, any of the 3 trained networks can correctly determine a previously unseen combination of these with a very high accuracy of about 93\% to 99\%. When it comes to determining links between a previously seen CVE and a brand new CVE, the results are much poorer. Although an average accuracy of roughly 80\% is achieved by all 3 networks, this is coupled with a poor F1 score around 60\%, contrasted with an F1 score of 96\% to 99\% for previously seen CVEs. Predictably, the accuracy and F1 scores for predicting links between completely new CVEs drops further, to levels which have questionable utility in a practical application at this stage.

This offers a variety of insights, as further analysis reveals that a bulk of the inaccuracies stems from the less represented (aka smaller) CWE groups, an issue also plaguing the work of Das et al., for which the authors successfully employ a "reconstruction decoder" to decrease the bias towards larger groups \cite{Das2021}. Employing regularization techniques such as this, alongside adjusting the selection of cliques and values of $p$ may yield more promising results when applied towards the new-to-new prediction problem. Interestingly, while Dataset 1 shows a positive correlation between size of CWE and accuracy, the other 2 experiments show a strong negative correlation. This could be due to the sampling mechanism, as the other 2 utilise random weighted sampling, but is more likely to do with the fact Dataset 1 contains more very small CWEs with a deceptively high accuracy, due to less of their positive examples appearing in the test set. F1 score remains positively correlated with CWE size in all experiments, which makes intuitive sense as more represented CWE groups are more likely to be remembered by the network.

\subsubsection{Fine-tuned vs base DistilBERT encoder}

In terms of comparison between the fine-tuned and base DistilBERT encoder, some remarkable results can be seen. Counter-intuitively, the fine-tuning process does not seem to have induced consistent improvement in performance, with the base encoder measurably outperforming the fine-tuned encoder in Dataset 3.

In Dataset 1, as seen in Table \ref{tab:clique_results_2}, when determining links in previously seen CVEs, the fine-tuned encoder only slightly outperforms the base DistilBERT encoder. However, a more dramatic improvement is seen when contrasting the old-to-new and new-to-new performance. In most cases, all metrics are boosted by a noticeable margin, which becomes more dramatic in new-to-new performance.

In Dataset 2, as seen in Table \ref{tab:random_results}, there is also a significant improvement when using the fine-tuned encoder, though the scale of this improvement is more visible in old-to-old experiments than in the Dataset 1 experiment. Measurable improvement is also seen in old-to-new and new-to-new prediction, though the F1 score remains poor.

Dataset 3, as seen in Table \ref{tab:random_results_2}, shows the most counter-intuitive results. In contrast to the previous 2 experiments, the base encoder shows slight improvement in predicting old-to-old and old-to-new CVE relationships, however this improvement becomes significant when predicting new-to-new CVEs.

Why this is the case remains rather unclear. Intuitively, this phenomenon could be explained by the fact that the prediction network might not use the more salient data obtained with the fine-tuned encoder when faced with CVEs it has previously examined, as it can learn more specific features of these CVEs (arguably, overfit to them), while when faced with unseen CVEs, it instead is forced to rely on more salient language features obtained using the fine-tuned encoder. However, this does not account for the fact that the fine-tuned encoder shows improvement in all data sampled but 1. This would most likely be explained by some other parameter of the data, such as the selection of CWEs and the language used within the description of its CVEs, or difference in the data distribution.

\subsubsection{Generation experiments and Consensus mechanism}

With respect to the experiments performed with Algorithm \ref{alg:generation_simple} and the consensus mechanism, it is clear that in all cases the consensus mechanism performs very well in minimising the impact of false positive and negative predictions on the stability of the generation algorithm. However, despite the networks achieving a significant level of accuracy in their predictions, with the average similarity level staying relatively high, there is still a notable amount of error, particularly when unseen CVEs are introduced. Results for Dataset 3 in Table \ref{tab:consensus} show much better results than in Table \ref{tab:consensus_old_new}, which may suggest that Dataset 3 experiments led to a degree of overfitting to seen CVEs to the detriment of generalizability. In contrast, Dataset 2 seems to have retained better results during generation, with the consensus similarity peaking at $77.5$ for CWEs 74 and 400, its notable that these are the largest CWEs in the dataset with $658$ and $1091$ CVEs respectively, compared to $253$ CVEs for the third largest CWE 326. Evidently, both the amount of underlying groups (CWEs) and the size of these groups present has a major effect on the results, with both datasets showing a notable decrease in similarity as one goes from 2 or 3 to 4 underlying groups being present in Table \ref{tab:consensus}, clearly showing that for tasks involving more than 2 or 3 groups a more robust approach is needed. A similar effect is seen in Table \ref{tab:consensus_old_new}, although not universally, as Dataset 3 boasts its 2nd largest consensus similarity for all 3 CWEs. Again, this is explained by the relative sizes of these, as CWE 203 has only $270$ CVEs compared to $2789$ and $700$ for CWEs 94 and 306. 

These datasets have been designed in this way deliberately, to study the degree to which smaller groups get lost in the midst of larger ones. While resource constraints meant that experiments could not be performed without negative link culling strategies, the estimated amount of negative links would have been $2322906$ for Dataset 2 and $6234292$. This is in stark contrast to the actual amount of negative links sourced using weighted random sampling, $779216$ and $2796013$ for each, a reduction of 3 and 2 fold respectively. Without this, negative links would have dominated the dataset and during initial experimentation on smaller datasets, the networks struggled to determine any positive links at all. As such, the experiments summarised in Tables \ref{tab:consensus} and \ref{tab:consensus_old_new} show that even without 2/3rds of negative links the networks can still learn to determine when CVEs are not related to each other.

\subsection{Practical viability}

Undoubtedly, the biggest limitation of the work presented thus far is the difficulty to meaningfully generalise to brand new CVEs. While not utilised in this work, approaches for better mitigating the imbalance of certain groups over others do exist, most relevant to this domain being the use of a reconstruction decoder by Das et al. \cite{Das2021}. It is clear from the results that including too many CWEs in the underlying training data can severely limit the performance and generality of the trained networks. In future work, methods proposed in this paper, such as the use of a consensus mechanism and the weighted random sampling of negative links should be combined with approaches such as the reconstruction decoder with a cohesive use-case or example to better determine the ability of this approach to generalise to new data.

Despite this limitation however, the prediction of old-to-old and old-to-new CVEs does have several practical applications. In cases where a series of empirical tests (e.g. penetration tests) has yielded groups of vulnerabilities that are observed to exist alongside one another, though without underlying contextualising information such as associated CWEs, this method could be used to train a model which can determine how these vulnerabilities could interact with new or hypothetical vulnerabilities based on text descriptions. In essence, the way in which CWEs are used to group together CVEs does not always align to observed co-existence of vulnerabilities, rather the sibling level relationships within CWE suggest that these CVEs are thought to induce or exploit similar weaknesses in various different systems. If instead, vulnerability groups were assigned based on observable dependency when performing an actual attack (put another way, exist under an OR node in an attack tree), these relationships could perhaps be learned and used to generate new attack trees in the manner described within this paper.

Another major result shown by the consensus experiment in Table \ref{tab:consensus} is that weighted random sampling of negative links does not seem to damage the generality of models trained on partial data, as the results were based on sampling groups of nodes for which there were 0 combinations shown in training data, yet a high degree of similarity for predicted groups was still achieved using Algorithm \ref{alg:generation_simple}. It is however unclear just how much the culling of negative links from otherwise legitimate examples adversely affects the end result, and more data would need to be collected to more effectively gauge the trade-off between performance and resource use.

While somewhat overlooked in this paper, the clique-based sampling method could potentially provide comparable performance on similar tasks, however the scope of this paper does not include formulating a domain specific case study robust enough to investigate this potential. A major limitation of the clique-based method is that the culling of negative links between arbitrarily defined groups of data can hurt the generalisation potential of the networks. While weighted random sampling does remove certain links, it never forbids the formation of links between specific groups of CWEs, with the definition of cliques introducing a level of bias that a random sample of all combinations does not. Thus, unless there is a justifiable reason for drawing such boundaries (such as one provided by a domain specific example), weighed random sampling appears to be the better, more generally applicable solution. Another approach could see a more general application of the concept behind the use of such cliques, or even a blend of clique and weighted random sampling, by defining CWE level associations on how relevant they are to one another (and thus if they should form negative links at a certain rate between one another), perhaps using the existing CWE hierarchy as a basis for this.

Therefore, while the weighted random sampling and consensus mechanism described in this paper provide a solid foundation, there is still a lot of work to be done before any such approach can be deployed towards a practically useful methodology for creating hierarchical attack models. More studies are needed to shed more light alongside additional solutions for tackling the main challenges towards practically viable cybersecurity activities, such as the construction of attack models based on empirical data.

\subsection{Future work}

Principally, this paper has set out to open up the possibility of utilising machine learning based predictive models for constructing hierarchical attack models in the style of attack trees. Although many questions have been addressed within this work, many remain to be investigated.

In terms of practical viability, work should be done on how viable vulnerability-to-vulnerability prediction is when using vulnerabilities grouped based on characteristics not described by CWE membership. As mentioned above, if a penetration test or an analysis of the data determined different groups of vulnerabilities, a method such as this should be deployed to determine how well a network can learn to associate future vulnerabilities within these groups.

In addition, more work needs to be done on the way positive and negative links are sampled between vulnerabilities, and a mechanism is needed to account for overlapping interpretations of group membership (e.g. vulnerabilities appearing in multiple groups) and underlying relationships in the data (e.g. CVEs belonging to CWEs that are linked to one another).

Also worth noting is the fact that beyond just CVE and CWE, there is a wide range of sources for hierarchical relationships between more abstract MITRE datasets, with there being at least 2 additional datasets placed above CWE \cite{hemberg2021linking}, which could provide a source for further development of the methods and algorithm introduced in this paper. While one might ask why existing work was not sufficient to build hierarchical threat models, such as the CVE to CWE prediction performed by Das et al.\cite{Das2021}, the focus in this paper was for a scheme driven purely by vulnerability data, with no contextualising data for vulnerability groups (e.g. CWE descriptions).

A major limitation of the method presented in this paper is the interpretation of negative links. That is, assuming that just because two vulnerabilities belong to different groups, they are not relevant to one another. This assumption is somewhat mitigated by the consensus mechanism, as the interpretation now goes to a consensus of multiple different measurements on the relevance of a given vulnerability, but there are still several issues. Namely, CVEs can feasibly be assigned under multiple CWEs, and the way in which negative links are sampled means that two CVEs may be assigned a negative link despite actually being relevant to one another. In essence, within this work there is no mechanism for resolving (rare though not trivial) cases with contradictory links, e.g. 2 CVEs being siblings in one CWE but not in another. In addition, CWEs follow a hierarchical structure with parent-child relationships of their own, which means that the assumption of no-relation can also be challenged from that angle, as a CVE in a child CWE and a CVE in a parent CWE may be argued to be related to one another after all. Notably, Das et al. factor in this hierarchical relationship in their work \cite{Das2021}, while this paper has focused on a more limited and direct implementation.

Immediately, the next step towards full generation of attack trees is a way to apply Algorithm \ref{alg:generation_simple} in a recursive layer-wise manner. This would consist of treating each newly generated set of parents as additional nodes for whom siblings must be assigned, thus the task would be to keep generating new layers of parent nodes until no more sibling relationships can be found. Conceptually, the main difficulty is in establishing how the embeddings would be derived for these new parent nodes, as the leaf nodes have their BERT embeddings sourced from text descriptions, a way to connect these to subsequent layers would need to be established. Another concern is the complexity of introducing additional layers, since predictably each new layer would have less and less nodes there would have to be some method for ensuring that these are not "lost" in the large amount of leaf nodes present in the training data. In the same vein, it should also hold that a subset of valid child nodes should result in embeddings similar to another subset, or the full set seen in training data.

This is a problem for which a wide variety of approaches could be applied. One approach could be to interpret the generation using Graph Neural Networks \cite{Scarselli2009}, as various tasks in the field of machine learning graph generation have found success with utilising learned representations for nodes and/or edges \cite{Faez2020}. A conceptually simpler approach would be to somehow utilise the text descriptions directly in the top layers, through methods such as concatenation of child descriptions and using summarisation techniques \cite{ATS_survey} to produce text descriptions for parents. An even simpler approach would be to simply take the average of the child embeddings, originating ultimately at the NLP embedded "leaf nodes" sourced from CVEs. It would also be necessary to incorporate logical labels for parent nodes in order to produce an attack tree generator \cite{pekaric2023streamlining,Schneider2000}, however how this could be achieved is also an open question, with Graph Neural Networks once again providing a possible way forward through learned representations of typed edges.

In the case of text encoders, future work must be done to investigate the impact of fine-tuning on the generalizability of trained networks, and fine tuning strategies beyond masked language modelling should be applied. Given the nature of the data, a transfer learning approach could be taken where a large subset of CVEs from one domain is used to train and fine-tune the initial encoder, and a smaller domain specific dataset is used to further apply the network for the domain. Further, work should be done on investigating how the full BERT encoder performs on this task, as due to resource limitations, this paper had to utilise the smaller DistilBERT model.

Finally, the ability to retain information on "rare" CVEs belonging to smaller CWE groups should be explored, most obviously by applying existing techniques such as reconstruction decoder by Das et al \cite{Das2021}. In cases where very few examples exist, the hierarchical nature of CWE may yield a larger set of training examples (from a larger "parent" CWE) on which to train an initial model, and then apply a sort of transfer learning to fine-tune the prediction network on the smaller set.

\section{Conclusion}

This paper has proposed a method for the prediction of sibling-level relationships between vulnerability data in the form of CVE entries grouped together using their associations to CWE entries. Having identified the challenges in terms of data complexity, particularly with data imbalance, two strategies for reducing class imbalance towards negative links have been outlined and empirically tested.

Additionally, an algorithm for using these predictions to building rudimentary groups of vulnerabilities was introduced alongside a consensus mechanism increasing the confidence of predicted links, which has been shown to be capable of producing groups that resemble the underlying CWE groupings for CVEs from which the training data was sampled. Finally, the remaining challenges and limitations of the work outlined in this paper were discussed together with the directions for future research needed to bridge the gap towards the full generation of hierarchical attack models using machine learning.

\section*{CRediT authorship contribution statement}
\textbf{Kacper Sowka}: Conceptualisation, Investigation, Data curation, Methodology, Software, Visualisation, Validation, Writing – original draft.
\textbf{Vasile Palade}: Supervision, Methodology, Conceptualisation, Writing – review \& editing, Resources.
\textbf{Hesam Jadidbonab}: Supervision, Conceptualisation, Writing – review \& editing.
\textbf{Xiaorui Jiang}: Supervision, Conceptualisation, Writing – review \& editing.

\section*{Declaration of competing interests}

The authors declare that beyond the co-funding of this research by Coventry University and HORIBA MIRA, as part of a PhD studentship, they have no known competing financial interests or personal relationships that could have appeared to influence the work reported in this paper. While the deliverables of the project (code, prototype, trained models etc) are the intellectual property of HORIBA MIRA held for the potential development of future services or products, this did not directly influence the contents of this paper.

\section*{Data availability}

While the code and trained models used to derive these results cannot be provided due to prior agreements in relation to funding, the data used was acquired from a publicly available instantiation of the BRON dataset and can be further distributed based on the MIT license which applies to BRON\footnote{https://github.com/ALFA-group/BRON/blob/master/LICENSE}. This selection can be found in JSON format using the DOI 10.17632/s2sw4ck42n.1.

\section*{Acknowledgements}

This research was co-funded by Coventry University and HORIBA MIRA.

\appendix

\section{Effect of different values of $p$ on weighted random sampling}
\label{sec:appendix_p}

For this Appendix, all CWEs acquired on 02/08/2022 with a cardinality larger than 2 have been compiled into a single dataset, which amounts to 220 CWES with the mean amount of CVE children being 555.93, the median being 58, mode being 2 and max being 19340. With respect to this data, the effect of different values for $p$ in Equation \ref{eq:sampling} can be seen in Table \ref{tab:pvalue}. Column "$p$" designates the value of $p$ used to derive the values for the given row, "\% positive links" refers to how many of the total links are positive, "\% from max" refers to the percentage of all children of the largest CWE group (with 19340 children) that were sampled for the given $p$ value, "\% from median" refers to the percentage of children sampled from a CWE with the median amount of children (58).

\begin{table}[!ht]\small 
    \centering
    \caption{Effect of different values of $p$ on the balance between positive and negative links}
    \begin{tabular}{|c|c|c|c|}
     \hline
     $p$ & \% positive links & \% from max & \% from median \\
     \hline
     1 & 6.2\% & 84.2\% & ~100\%\\
     \hline
     2 & 6.9\% & 70.9\% & 99.9\%\\
     \hline
     3 & 7.6\% & 59.7\% & 99.9\%\\
     \hline
     4 & 8.3\% & 50.2\% & 99.8\%\\ 
     \hline
     5 & 9.0\% & 42.3\% & 99.8\%\\
     \hline
     6 & 9.8\% & 35.6\% & 99.7\%\\
     \hline
     7 & 10.5\% & 30.0\% & 99.7\%\\
     \hline
     8 & 11.3\% & 25.2\% & 99.6\%\\
     \hline
     9 & 12.1\% & 21.2\% & 99.6\%\\
     \hline
     10 & 12.9\% & 17.9\% & 99.5\%\\
     \hline
    \end{tabular}
    \label{tab:pvalue}
\end{table}

\section{Fine-tuning on CVE data using Masked Language Modelling}
\label{sec:appendix_MLM}

When fine tuning the DistilBERT encoder, all CVE data acquired from BRON on 02/08/2022 was collected, with each CVE description being tokenized using the Huggingface DistilBERT tokenizer, which was used with the uncased Huggingface DistilBERT model\footnote{https://huggingface.co/distilbert-base-uncased}. The following hyperparameters were used during the fine-tuning process:

\begin{itemize}
    \item Max sequence length: 128
    \item Data entries: 175294
    \item Train split: 75\% (131930)
    \item Validation split: 10\% (16978)
    \item Test split: 15\% (26386)
    \item Epochs: 4
    \item Warmup steps: 500
    \item Loss: Cross entropy across logits for each word in the vocabulary
\end{itemize}

\begin{table*}[h]
    \centering
    \caption{Results showing empirical performance of fine tuning on distilbert with respect to batch size and learning rate on all CVE data}
    \begin{tabular}{|c|l|l|l|l|l|c|c|}
    \hline
        \textbf{Batch Size} & \textbf{Learning rate} & \textbf{Loss} & \textbf{Exact} & \textbf{Top 5} & \textbf{Incorrect} & \textbf{Exact} & \textbf{Top 5}  \\ \hline
         & 2.00E-05 & 0.1363 & 129122 & 161711 & 52139 & 60\% & 76\% \\ \cline{2-8}
         \textbf{16} & 3.00E-05 & 0.1291 & 132928 & 164595 & 49255 & 62\% & 77\% \\ \cline{2-8}
         & 5.00E-05 & 0.1212 & 137031 & 167819 & 46031 & 64\% & 78\% \\ \hline
         & 2.00E-05 & 0.1507 & 121668 & 155974 & 92182 & 57\% & 73\% \\ \cline{2-8}
         \textbf{32} & 3.00E-05 & 0.1415 & 126397 & 159648 & 54202 & 59\% & 75\% \\ \cline{2-8}
         & 5.00E-05 & 0.1313 & 131685 & 163694 & 50156 & 62\% & 77\% \\ \hline
        \multicolumn{2}{|c|}{\textbf{distilbert-base-uncased}} & 9.9554 & 56564 & 91387 & 122463 & 26\% & 43\% \\ \hline
    \end{tabular}
    \label{tab:distil_fine}
\end{table*}
\begin{table*}[h]
    \centering
    \caption{Comparison between the best performing masked language model when tested on two different datasets}
    \begin{tabular}{|l|l|l|l|l|c|c|}
    \hline
        \textbf{Set} & \textbf{Loss} & \textbf{Exact} & \textbf{Top 5} & \textbf{Incorrect} & \textbf{Accuracy} & \textbf{Top 5 Accuracy}  \\ \hline
        Test & 0.1218 & 87587 & 107170 & 29899 & 64\% & 78\% \\ \hline
        Validation & 0.1212 & 137031 & 167819 & 46031 & 64\% & 78\% \\ \hline
    \end{tabular}
    \label{tab:distil_fine_comparison}
\end{table*}

Table \ref{tab:distil_fine} shows the results for all the hyperparameters suggested for fine-tuning in the BERT paper \cite{Devlin2019}, including masking 15\% of all tokens, with each column having the following meaning:

\begin{itemize}
    \item Loss: Cross entropy loss between the target logits and the predicted logits for each word in the vocabulary
    \item Exact: Amount of times the model predicted the correct word exactly (most likely prediction was the actual masked word as per the label)
    \item Top 5: Amount of times the top 5 predictions by the model included the actual masked word (includes exact predictions)
    \item Incorrect: Amount of incorrect guesses made by the model (not in the top 5)
    \item Exact accuracy: Percentage of times that the predicted word was guessed exactly correct $\frac{exact}{incorrect+top5}$
    \item Top 5 accuracy: Percentage of times that the top 5 predicted words included the actual masked word $\frac{top5}{incorrect+top5}$
\end{itemize}

As the best performing model, the DistilBERT model fine-tuned using a batch size of $16$ and a learning rate of $5.00E-05$ was chosen as the fine-tuned encoder for every experiment. A further run on these hyperparameters using the test set produces remarkably similar results, seen compared with the earlier validation results in Table \ref{tab:distil_fine_comparison}, which solidifies the generality of the fine-tuned model.

\section{Dataset}
\label{sec:data_appendix}
For all experiments performed in this paper, data on the text descriptions of CVEs and their links to CWEs were sourced using the publicly available ArangoDB BRON database\footnote{http://bron.alfa.csail.mit.edu/info.html}, collected using the ArangoDB API. An interesting quirk of this with relation to the data available for the experiments used, was that when data was first collected on 02/08/2022 there were significantly less CVE connections available than if the data was fetched now, by a factor of over a 100, and many of the "new" connections are questionable. Naturally, this means that the dataset used in this paper is a subset of the "full" data that is available, which was always going to be the case given the staggering amount of data available via CVE alone. Principally, the decision to work with this "reduced" dataset was to allow for experiments involving a large amount of possible groups, and the 100-fold increase in connections is suspicious, as many of them seem spurious. On the latter point, comparing the CVE to CWE connections in the "older" dataset and what can be derived by fetching data from the NVD API for CVEs\footnote{https://services.nvd.nist.gov/rest/json/cves/1.0} and the "Observed Examples" section of CWE data\footnote{https://cwe.mitre.org/data/csv/1000.csv.zip}. When performing a rudimentary comparison between the links present in the "original" repositories and the ones present in BRON, for the "older" BRON collection $115929$ out of $119745$ links are exactly the same ($96.8\%$), while in a "new" version of BRON collected on 18/08/2023 only $1571$ out of $177438$ ($0.9\%$) are exactly the same, with the average increase in links being $23.39$. Therefore, given this huge increase in links, it is difficult to verify the veracity of these new links and so an older version more in line with the original data is being used.

A copy of this data subset can be found in JSON format using the DOI "10.17632/s2sw4ck42n.1". It is important to remember that the edges referred to in the BRON data are not analogous to the relationships predicted in this paper, as the BRON data considers parent-child relations while this paper is concerned with siblings, which can be derived form parent-child relations.

Importantly, the choice of the CVE dataset was as a generic representative for vulnerability datasets frequently receiving new entries rather than the only relevant use case, the intention of this work was to demonstrate the overarching principles behind using neural networks, NLP and the techniques outlined to manage cybersecurity data. In principle, the idea is that by using the current understanding of vulnerabilities and their relationship to each other you can train neural networks to learn underlying patterns in cases where there is no data for what characterises the groups of vulnerabilities, allowing for quicker vulnerability analysis.

New CVEs were acquired on $06/10/2023$ using the above described NVD and CWE APIs.

For each dataset, the following selection of CWEs was used, with the CVE cardinality being shown in square brackets.
\subsection{Dataset 1}
Numbers designate the "cliques" of CWEs in Dataset 1.
\begin{enumerate}
    \item CWE 497: Exposure of Sensitive System Information to an Unauthorized Control Sphere [3], CWE 326: Inadequate Encryption Strength [237], CWE 613: Insufficient Session Expiration [163], CWE 1284: Improper Validation of Specified Quantity in Input [2], CWE 61: UNIX Symbolic Link (Symlink) Following [8].
    \item CWE 122: Heap-based Buffer Overflow [84], CWE 281: Improper Preservation of Permissions [107], CWE 772: Missing Release of Resource after Effective Lifetime [370], CWE 943: Improper Neutralization of Special Elements in Data Query Logic [3], CWE 426: Untrusted Search Path [373].
    \item CWE 732: Incorrect Permission Assignment for Critical Resource [865], CWE 303: Incorrect Implementation of Authentication Algorithm [3], CWE 228: Improper Handling of Syntactically Invalid Structure [2], CWE 538: Insertion of Sensitive Information into Externally-Accessible File or Directory [10], CWE 763: Release of Invalid Pointer or Reference [39].
\end{enumerate}

\subsection{Dataset 2}

CWE 345: Insufficient Verification of Data Authenticity [216], CWE 693: Protection Mechanism Failure [45], CWE 294: Authentication Bypass by Capture-replay [76], CWE 74: Improper Neutralization of Special Elements in Output Used by a Downstream Component ('Injection') [658], CWE 288: Authentication Bypass Using an Alternate Path or Channel [12], CWE 326: Inadequate Encryption Strength [253], CWE 400: Uncontrolled Resource Consumption [1091], CWE 754: Improper Check for Unusual or Exceptional Conditions [183]

\subsection{Dataset 3}

CWE 682: Incorrect Calculation [69], CWE 697: Incorrect Comparison [44], CWE 306: Missing Authentication for Critical Function [700], CWE 665: Improper Initialization [203], CWE 94: Improper Control of Generation of Code ('Code Injection') [2789], CWE 824: Access of Uninitialized Pointer [109], CWE 203: Observable Discrepancy [270], CWE 770: Allocation of Resources Without Limits or Throttling [402]

 \bibliographystyle{elsarticle-num} 
 \bibliography{cas-refs}





% \end{thebibliography}
\end{document}